\documentclass[runningheads]{llncs}
\usepackage{multirow}
\usepackage{graphicx}
\usepackage{algorithm,algcompatible,amsmath,amssymb}
\usepackage{hyperref}
\DeclareMathOperator*{\argmin}{argmin} 
\algnewcommand\INPUT{\item[\textbf{Input:}]}%
\algnewcommand\OUTPUT{\item[\textbf{Output:}]}%
\renewcommand{\COMMENT}[2][.6\linewidth]{\leavevmode\hfill\makebox[#1][l]{\#~#2}}
\usepackage[colorinlistoftodos]{todonotes}
\usepackage{booktabs}
\usepackage{bbding}

\begin{document}
\title{Deep Generative Model-based Quality Control for Cardiac MRI Segmentation}
\titlerunning{Deep Generative Model-based Quality Control}
%

\author{Shuo Wang\inst{1}(\Envelope), Giacomo Tarroni \inst{2},  Chen Qin \inst{2},\\ Yuanhan Mo \inst{1}, Chengliang Dai  \inst{1}, Chen Chen \inst{2}, \\ Ben Glocker \inst{2},
Yike Guo\inst{1}, Daniel Rueckert\inst{2}, and Wenjia Bai\inst{1,3}}
\authorrunning{S. Wang et al.}
\institute{Data Science Institute, Imperial College London, London, UK \\ 
\and BioMedIA Group, Department of Computing, Imperial College London, UK \\ \and Department of Brain Sciences, Imperial College London, UK
\email{shuo.wang@imperial.ac.uk}}


\maketitle              
\begin{abstract}
In recent years, convolutional neural networks have demonstrated promising performance in a variety of medical image segmentation tasks. However, when a trained segmentation model is deployed into the real clinical world, the model may not perform optimally. A major challenge is the potential poor-quality segmentations generated due to degraded image quality or domain shift issues. There is a timely need to develop an automated quality control method that can detect poor segmentations and feedback to clinicians. Here we propose a novel deep generative model-based framework for quality control of cardiac MRI segmentation. It first learns a manifold of good-quality image-segmentation pairs using a generative model. The quality of a given test segmentation is then assessed by evaluating the difference from its projection onto the good-quality manifold. In particular, the projection is refined through iterative search in the latent space. The proposed method achieves high prediction accuracy on two publicly available cardiac MRI datasets. Moreover, it shows better generalisation ability than traditional regression-based methods. Our approach provides a real-time and model-agnostic quality control for cardiac MRI segmentation, which has the potential to be integrated into clinical image analysis workflows.

\end{abstract}
\section{Introduction}
Cardiovascular diseases (CVDs) are the leading cause of death globally, taking more than 18 million lives every year \cite{WHOScaleStroke.}. Cardiac magnetic resonance imaging (MRI) has been widely used in clinical practice for evaluating cardiac structure and function. To derive quantitative measures from cardiac MRI, accurate segmentation is of great importance. Over the past few years, various architectures of convolutional neural networks (CNNs) have been developed to deliver state-of-the-art performance in the task of automated cardiac MRI segmentation \cite{bai2018automated,tao2019deep,bernard2018deep,zheng20183}. Although satisfactory performance has been achieved on specific datasets, care must be taken when deploying these models into clinical practice. In fact, it is inevitable for automated segmentation algorithms (not limited to CNN-based) to generate a number of poor-quality segmentations in real-world scenarios, due to differences in scanner models and acquisition protocols as well as potential poor image quality and motion artifacts. Therefore, reliable quality control (QC) of cardiac MRI segmentation on a per-case basis is highly desired and of great importance for successful translation into clinical practice.

\subsubsection{Related work:} Numerous efforts have been devoted into quality control of medical images \cite{tarroni2018learning,carapella2016towards,zhang2016automated} and segmentations \cite{robinson2019automated,alba2018automatic}. In this work, we focus on the latter, i.e. segmentation quality control. Existing literature can be broadly classified into two categories:

\noindent\textbf{Learning-based quality control:} These methods consider quality control as a regression or classification task where a quality metric is predicted from extracted features. \cite{kohlberger2012evaluating} proposed 42 hand-crafted features based on intensity and appearance and achieved an accuracy of 85\% in detecting segmentation failure. \cite{robinson2018real} developed a CNN-based method for real-time regression of the Dice similarity metric from image-segmentation pairs. \cite{hann2019quality} integrated quality control into the segmentation network by regressing the Dice metric. Most of these methods require poor-quality segmentations as negative samples to train the regression or classification model. This makes quality control specific to the segmentation model and the type of poor-quality segmentations used for training. \cite{liu2019alarm} used a variational auto-encoder (VAE) for learning the shape features of segmentation in an unsupervised manner and proposed to use the evidence lower bound (ELBO) as a predictor. This model-agnostic structure provides valuable insights and an elegant theoretical framework for quality control.

\noindent\textbf{Registration-based quality control:} These methods perform image registration between the test image with a set of pre-selected template images with known segmentations. Then the quality metric can be evaluated by referring to the warped segmentations of these template images. Following this direction, \cite{valindria2017reverse} proposed the concept of reverse classification accuracy (RCA) to predict segmentation quality and \cite{robinson2019automated} achieved good performance on a large-scale cardiac MRI dataset. These methods can be computationally expensive due to the cost of multiple image registrations, which could potentially be reduced by using GPU acceleration and learning-based registration tools \cite{Haskins2020DeepSurvey}. 

\subsubsection{Contributions:} 
There are three major contributions of this work. Firstly, we propose a generic deep generative model-based framework which learns the manifold of good-quality segmentations for quality control on a per-case basis. Secondly, we implement the framework with a VAE and propose an iterative search strategy in the latent space. Finally, we compare the performance of our method with regression-based methods on two different datasets, demonstrating both the accuracy and generalisation ability of the method.


\section{Methodology}
\subsection{Problem Formulation} Let $F$ denote an arbitrary type of segmentation model to be deployed. Given a test image $I$, the segmentation model provides a predicted segmentation $\hat{S}=F(I)$. The ground-truth quality of $\hat{S}$ is defined as $q(S_{gt}, \hat{S})$ where $S_{gt}$ is the ground-truth segmentation and $q$ is a chosen quality metric (e.g. Dice metric). The aim of quality control is to develop a model $Q$ so that $Q( \hat{S};I)\approx q(S_{gt}, \hat{S})$.

\subsection{Deep Generative Model-based Quality Control}
\begin{figure}
\centering
\includegraphics[width=\textwidth]{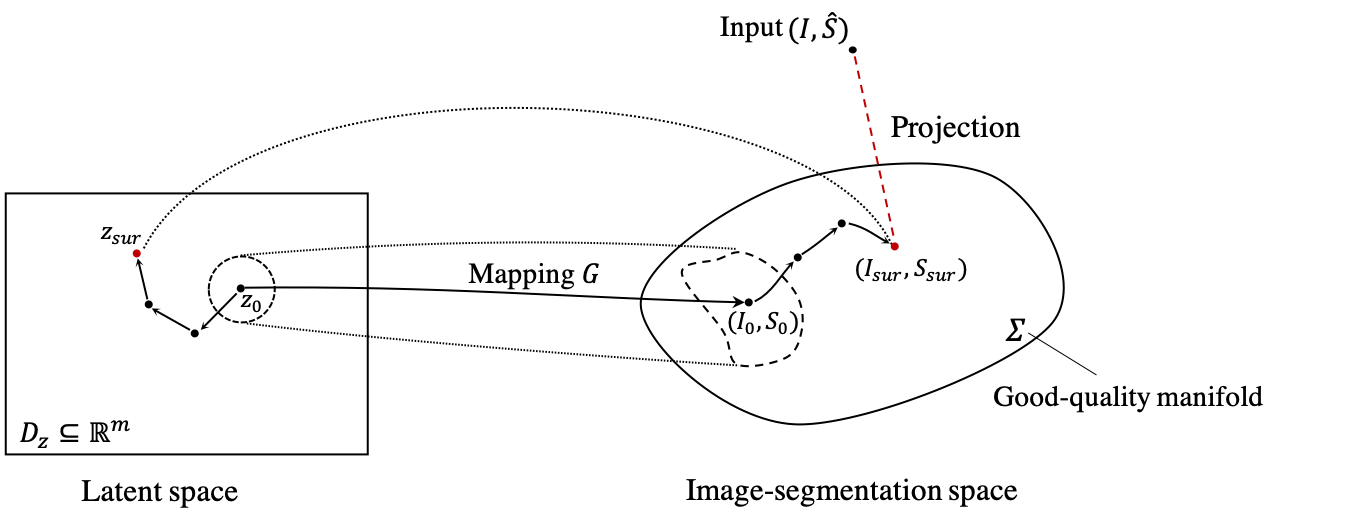}
\caption{Overview of the deep generative model-based quality control framework. The generative model $G$ is trained to learn a mapping $G(z)$ from the low-dimensional latent space $D_{z}$ to the good-quality manifold $\Sigma$. The input image-segmentation pair $(I, \hat{S})$ is projected to $(S_{sur}, I_{sur})$ on the manifold through iterative search, which is in turn used as surrogate ground truth for quality prediction. $z_0$ is the initial guess in the latent space and it converges to $z_{sur}$. } \label{figure1}
\end{figure}

Quality control would be trivial if the ground-truth segmentation $S_{gt}$ was available. Intuitively, the proposed framework aims to find a good-quality segmentation $S_{sur}$ as a surrogate for ground truth so that $q(S_{sur}, \hat{S}) \approx q(S_{gt}, \hat{S})$. This is realised through iterative search on the manifold of good-quality segmentations (Fig.\ref{figure1}).

\subsubsection{Good-quality manifold:}
The core component of this framework is a deep generative model $G$ which learns how to generate good-quality image-segmentation pairs. Formally, let $X=(I, S)\in D_I \times	D_S$ represent an image-segmentation pair, where $D_I$ and $D_S$ are the domains of images and possible segmentations. The key assumption of this framework is that good-quality pairs $(I, S_{gt})$ are distributed on a manifold $\Sigma \subset D_I \times D_S$, named as \textit{good-quality manifold}. The generator $G$ learns to construct a low-dimensional latent space $D_z \subseteq \mathbb{R}^m$ and a mapping to the good-quality manifold:
\begin{equation}
G(z): D_z \ni z \mapsto X=G(z) \in D_I \times D_S
\end{equation}
where $z$ denotes the latent variable with dimension $m$. The mapping $G(z)$ is usually intractable but can be approximated using generative models such as generative adversarial networks (GANs) or VAEs. 

\subsubsection{Iterative search in the latent space:} To incorporate the generator into the quality control framework, we develop an iterative search scheme in the latent space to find a surrogate segmentation for a given image-segmentation pair as input. This surrogate segmentation is used for quality prediction. Finding the closest surrogate segmentation (i.e. projection) on the good-quality manifold is formulated as an optimisation problem,
\begin{equation}
z_{sur} = \argmin_{z \in D_z} \mathcal{L}(G(z), (I,\hat{S}))
\end{equation}
which minimises the distance metric $\mathcal{L}$ between the reconstructed $G(z)$ and the input image-segmentation pair $(I,\hat{S})$. This problem can be solved using the gradient descent method as explained in Algorithm \ref{iter_scheme}.

\begin{algorithm}
    \caption{Iterative search of surrogate segmentation for quality prediction}\label{iter_scheme}
  \begin{algorithmic}[1]
    \REQUIRE A trained generator $G: D_z \ni z \mapsto G(z) = (I,S) \in \Sigma$
    \INPUT Image-segmentation pair $(I,\hat{S})$
    \OUTPUT Quality prediction $Q(\hat{S};I)$
    \STATE \textbf{Initialization} $z=z_0 \in D_z$
    \WHILE{$L$ not converge}
        \STATE $L = \mathcal{L}(G(z),(I,\hat{S}))$
        \STATE $grad = \nabla_{z}L$ \COMMENT{calculate gradient through back-propagation}
        \STATE $z = z - \alpha \cdot grad$ \COMMENT{gradient descent with learning rate $\alpha$}
    \ENDWHILE
    \STATE $S_{sur}=G(z_{sur})$ 
    \STATE $Q(\hat{S};I)=q(S_{sur},\hat{S})$
    \COMMENT{perform quality control $q$ (e.g. Dice)}
  \end{algorithmic}
\end{algorithm}

\subsection{Generative Model using VAE} 
\begin{figure}
\includegraphics[width=\textwidth]{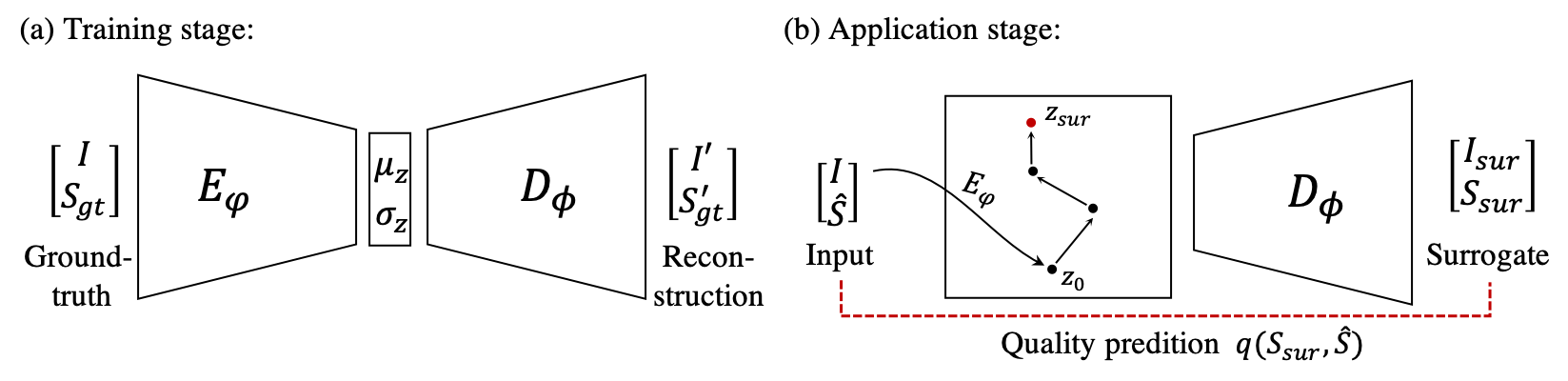}
\caption{Framework implementation using the variational autoencoder (VAE). In the training stage, the ground-truth image-segmentation pairs are used. In the application stage, the VAE decoder is used as the generator for iterative search of the surrogate segmentation on the good-quality manifold. Initial guess $z_0$ is from the encoder. 
} \label{figure2}
\end{figure}

The proposed framework can be implemented with different generative models as long as a good-quality segmentation generator with smooth latent space is available. In this paper, we employ the VAE (Fig \ref{figure2}) which includes an encoder $E_\varphi$ and a decoder $D_\phi$, where $\varphi$ and $\phi$ denote the model parameters \cite{kingma2013auto}. The image-segmentation pair $(I, S)$ is encoded by $E_{\varphi}$ to follow a Gaussian distribution $\mathcal{N}(\mu_z, \sigma_z^2)$ in the latent space, where $\mu_z$ and $\sigma_z^2$ denote the mean and variance respectively. A probabilistic reconstruction of the image-segmentation pair $(I', S')$ is generated from the decoder $D_\phi$.

At the training stage, the ground-truth image-segmentation pairs are used to train the VAE. The loss function includes a reconstruction loss and a KL divergence term for regularisation \cite{higgins2017beta}:
\begin{equation}
    \mathcal{L}_{VAE}=\mathcal{L}_{recon}+\beta \cdot D_{KL}(\mathcal{N}(\mu_z, \sigma_z^2)||\mathcal{N}(0, \textbf{I}))
\end{equation}
\begin{equation}
    \mathcal{L}_{recon}=BCE(S_{GT}, S_{GT}') + MSE(I, I')
\end{equation}
where $\beta$ is the hyperparameter that balances between reconstruction loss and regularisation. The reconstruction loss is evaluated using the binary cross-entropy (BCE) for segmentation and the mean square error (MSE) for image, respectively. The effects of the weight $\beta$ and the latent space dimension $m$ will be evaluated in the ablation study. 

At the application stage, the VAE decoder $D_\phi$ is used as the generator, reconstructing image-segmentation pairs from the latent space. The initial guess $z_0$ in the latent space is obtained from the encoder $E_{\varphi}$.
Following Algorithm \ref{iter_scheme}, the surrogate segmentation $S_{sur}$ can be found via iterative search (Fig. \ref{figure2}b). Finally, the quality metric is evaluated by $q(S_{sur},\hat{S})$, e.g. Dice metric.

\section{Experiments}

\subsection{Datasets}
\noindent \textbf{UK Biobank dataset:}
Short-axis cardiac images at the end-diastolic (ED) frame of 1,500 subjects were obtained from UK Biobank and split into three subsets for training (800 cases), validation (200 cases) and test (500 cases). The in-plane resolution is 1.8x1.8 mm with slice thickness of 8 mm and slice gap of 2 mm. A short-axis image stack typically consists of 10 to 12 image slices. Ground-truth good segmentations were generated from a publicly available fully-convolutional network (FCN) that has demonstrated a high performance \cite{bai2018automated}, with manual quality control by an experienced cardiologist.

\noindent \textbf{ACDC dataset:}
100 subjects including a normal group and four pathology groups were obtained from ACDC dataset \cite{bernard2018deep} and resampled to the same spatial resolution as UK Biobank data. The ground-truth segmentations at the ED frame were provided by the ACDC challenge organisers.

\subsection{Experimental Design}
In this study, we evaluate the performance of our proposed method and compare with two regression-based methods for quality control of cardiac MRI segmentation. Specifically, we focus on the myocardium, which is a challenging cardiac structure to segment and of high clinical relevance.

\subsubsection{VAE implementation and training:}\label{implementation}
The VAE encoder is composed of four convolutional layers (each was followed by $ReLU$ activation), and one fully connected layer. The decoder has a similar structure with reversed order and the last layer is followed by $Sigmoid$ function. The latent space dimension $m$ was set to 8 and the hyperparameter $\beta$ was set to 0.01 from ablation study results. The architecture is shown in Appendix Fig. A1. The model was implemented in PyTorch and trained using the Adam optimiser with learning rate 0.0001 and batch size 16. It was trained for 100 epochs and an early stopping criterion was used based on the validation set performance. To improve the computational efficiency, the VAE was trained on a region of interest (ROI) centered around the myocardium with the window size of 96x96 pixel, which was heuristically determined to include the whole cardiac structure. The cropped image intensity was normalised to the $[0, 1]$ range and stacked with the binary segmentation. 

\subsubsection{Baseline methods:} Two regression-based methods were used as baselines: a) a support vector regression (SVR) model with 42 hand-crafted features about shape and appearance \cite{kohlberger2012evaluating} and b) a CNN regression network (ResNet-18 backbone) with the image-segmentation pair as input \cite{robinson2018real}. Both baseline methods use the Dice metric as the regression target.

\subsubsection{Experiment 1: UK Biobank} Besides the ground-truth segmentations, we generated poor-quality segmentations by attacking the segmentation model. White noise with different variance level was added to the original images, resulting in a dataset of poor-quality segmentations with uniform Dice distribution. The quality prediction was performed on the test set of the attacked segmentations.

\noindent \textbf{Experiment 2: ACDC} We deployed a UK Biobank trained segmentation model on ACDC dataset without fine-tuning. This reflects a real-world clinical setting, where segmentation failures would occur due to domain shift issues. 

\noindent \textbf{Ablation study:} We adjusted the dimensionality of the latent space $m$ and the hyperparameter $\beta$ and performed a sensitivity analysis on the UK Biobank validation dataset. The result is reported in the Appendix Table A1.  

\section{Results and Discussion}
\begin{figure} 
\centering
\includegraphics[width=300pt]{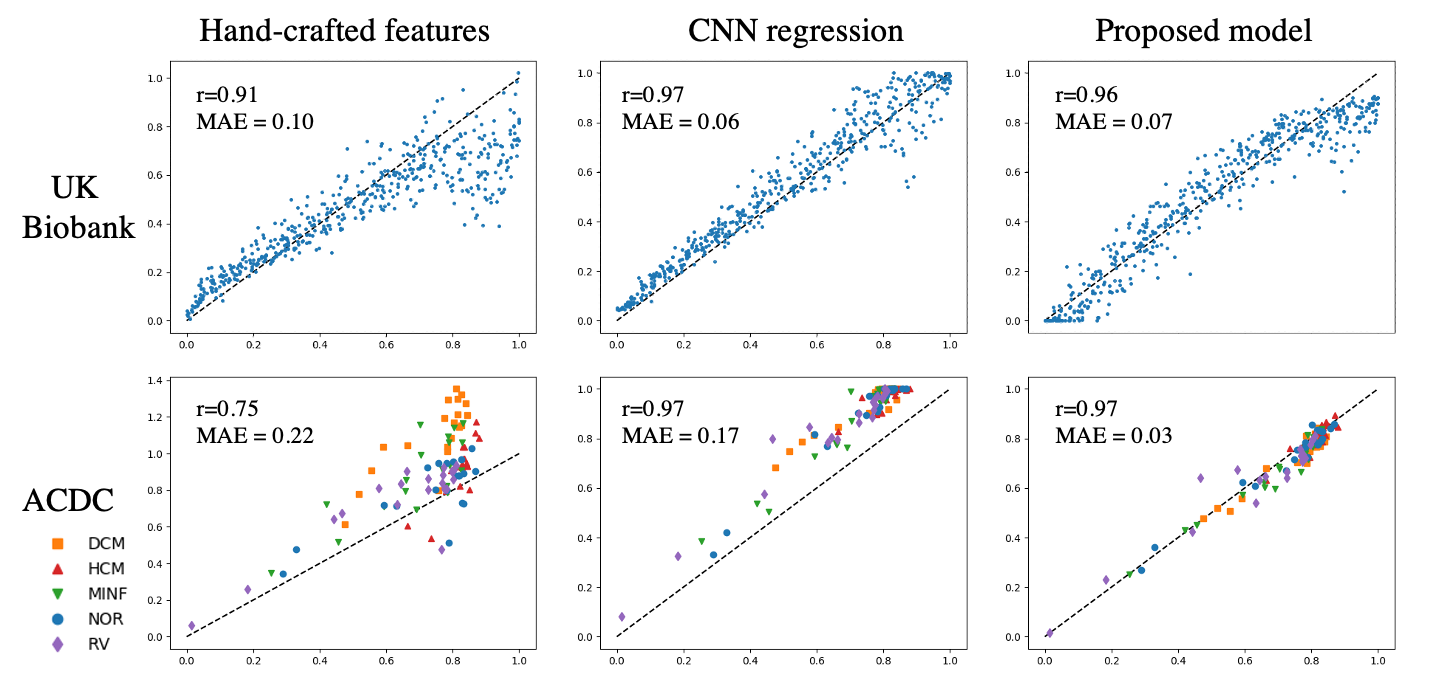}
\caption{Comparison of the performance of different quality control methods. The x-axis is the real Dice of each subject and the y-axis is the predicted Dice by each method. The dashed line is the $y=x$, plotted for reference. Top row: UK Biobank data ($n=500$). Bottom row: ACDC data ($n=100$), with five subgroups plotted in different colors.
} \label{figure3}
\end{figure}

Quality control performance is assessed in terms of Dice metric prediction accuracy. Pearson correlation coefficient $r$ and mean absolute error (MAE) between predicted Dice and real Dice are calculated. Table 1 compares the quantitative performance of the methods and Fig.\ref{figure3} visualises the predictions. On UK Biobank dataset, the hand-crafted feature method performed the worst. The proposed method achieved a similar performance ($r$=0.96, MAE=0.07) as the CNN regression method ($r$=0.97, MAE=0.06). However, on ACDC dataset, the proposed method ($r$=0.97, MAE=0.03) outperformed the CNN regression method ($r$=0.97, MAE=0.17) with a smaller MAE. As shown in Fig. \ref{figure3}, on ACDC dataset, the prediction of the proposed method aligns well with the identity line, whereas the CNN regression method clearly deviates from the line, even though the $r$ coefficient is still high.

A possible explanation for this is that the proposed generative method works by learning the good-quality manifold and proposing the surrogate ground truth for quality assessment. Thus it is agnostic to the types of poor-quality segmentations. The training of hand-crafted features and CNN-based regression methods require poor-quality segmentations and may be overfitted to the UK Biobank data. When they are deployed onto ACDC dataset, there is a shift not only for image appearance (e.g. difference between 1.5T and 3.0T MRI scanner) but also for types of segmentation failures. In addition, the ACDC dataset consists of more pathological cases, whereas the UK Biobank comes from a general healthy population. Due to the domain shift, the performance of regression-based methods degraded. In contrast, the proposed method maintained a high prediction accuracy against domain shift. This indicates the advantage of a generative model-based framework for generalisation. It also can be potentially used as a system to monitor the performance of deployed segmentation models over time.

\begin{table}[]
\label{table1}
\caption{Quality control performance of three models on two cardiac datasets. The Pearson correlation coefficient $r$ and the mean absolute error (MAE) between predicted and true Dice metrics are reported. MAE is reported as mean (standard deviation). For ACDC dataset, the performance on five subgroups \cite{bernard2018deep} are aslo reported.}
\centering
\resizebox{\textwidth}{!}{%
\begin{tabular}{llcccccc}
\toprule
\multicolumn{2}{c}{\multirow{2}{*}{Dataset\ \ }} & \multicolumn{2}{c}{Hand-crafted features \cite{kohlberger2012evaluating}} & \multicolumn{2}{c}{CNN regression \cite{robinson2018real}} & \multicolumn{2}{c}{Proposed model} \\ \cline{3-8} 
\multicolumn{2}{c}{} & \multicolumn{1}{c}{$r$} & MAE & \multicolumn{1}{c}{$r$} & MAE & \multicolumn{1}{c}{$r$} & MAE \\ \midrule
\multicolumn{2}{l}{UK Biobank} & 0.909 & 0.100(0.100) & \textbf{0.973} & \textbf{0.061(0.049)} & 0.958 & 0.067(0.052) \\ \midrule
\multicolumn{2}{l}{ACDC} & 0.728 & 0.182(0.130) & 0.968 & 0.165(0.044) & \textbf{0.969} & \textbf{0.033(0.028)} \\ \midrule
 & \ DCM & 0.802 & 0.353(0.123) & 0.956 & 0.186(0.034) & \textbf{0.964} & \textbf{0.036(0.020)} \\ 
 & \ HCM & 0.838 & 0.131(0.075) & 0.836 & 0.155(0.027) & \textbf{0.896} & \textbf{0.023(0.015)} \\ 
 & \ MINF & 0.815 & 0.184(0.123) & 0.969 & 0.156(0.051) & \textbf{0.976} & \textbf{0.033(0.029)} \\ 
 & \ NOR & 0.775 & 0.114(0.065) & 0.985 & 0.158(0.040) & \textbf{0.985} & \textbf{0.026(0.017)} \\ 
 & \ RV & 0.877 & 0.129(0.073) & 0.974 & 0.169(0.053) & \textbf{0.960} & \textbf{0.045(0.042)} \\ \bottomrule
\end{tabular}%
}
\end{table}

\begin{figure} 
\centering
\includegraphics[width=280pt]{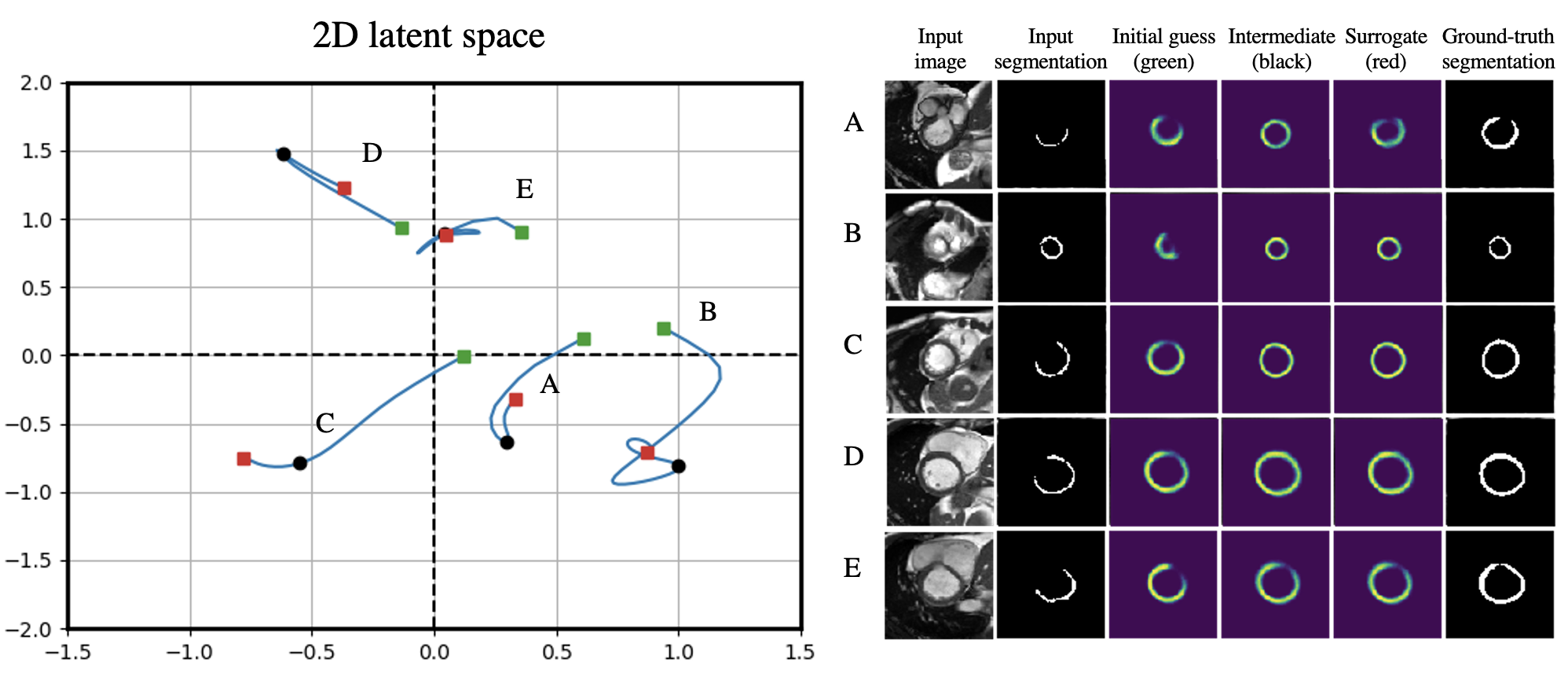}
\caption{Visualisation of searching path in a two-dimensional latent space. Left: searching paths for five exemplar samples (green point: initial guess from the VAE encoder; black point: intermediate state during iterative search; red point: convergence point for surrogate segmentation). Right: the input image and segmentation, reconstructed segmentations along the searching path and the ground-truth segmentation.} \label{figure4}
\end{figure}

To gain insights into our proposed method, we visualised several searching paths within a two-dimensional latent space and corresponding image-segmentation pairs reconstructed by our generative model (Fig. \ref{figure4}). The poor-quality segmentation could be projected onto the good-quality manifold and refined iteratively to obtain the surrogate segmentation. The surrogate segmentation on the good-quality manifold is more plausible and can potentially be used as $prior$ to correct poor-quality segmentations. It is also expected that the performance could be improved using advanced generative models and better manifold learning \cite{bojanowski2017optimizing}.

\section{Conclusion}
Here we propose a generative-model based framework for cardiac image segmentation quality control. It is model-agnostic, in the sense that it does not depend on specific segmentation models or types of segmentation failures. It can be potentially extended for quality control for different anatomical structures. 

\newpage

%
%
\bibliographystyle{splncs}
\bibliography{ref}

\newpage
\pagestyle{empty}
\setcounter{secnumdepth}{0}
\section{Appendix}

\setcounter{figure}{0}
\setcounter{table}{0}
\renewcommand{\thefigure}{A\arabic{figure}}
\renewcommand{\thetable}{A\arabic{table}}

\begin{figure} 
\includegraphics[width=\textwidth]{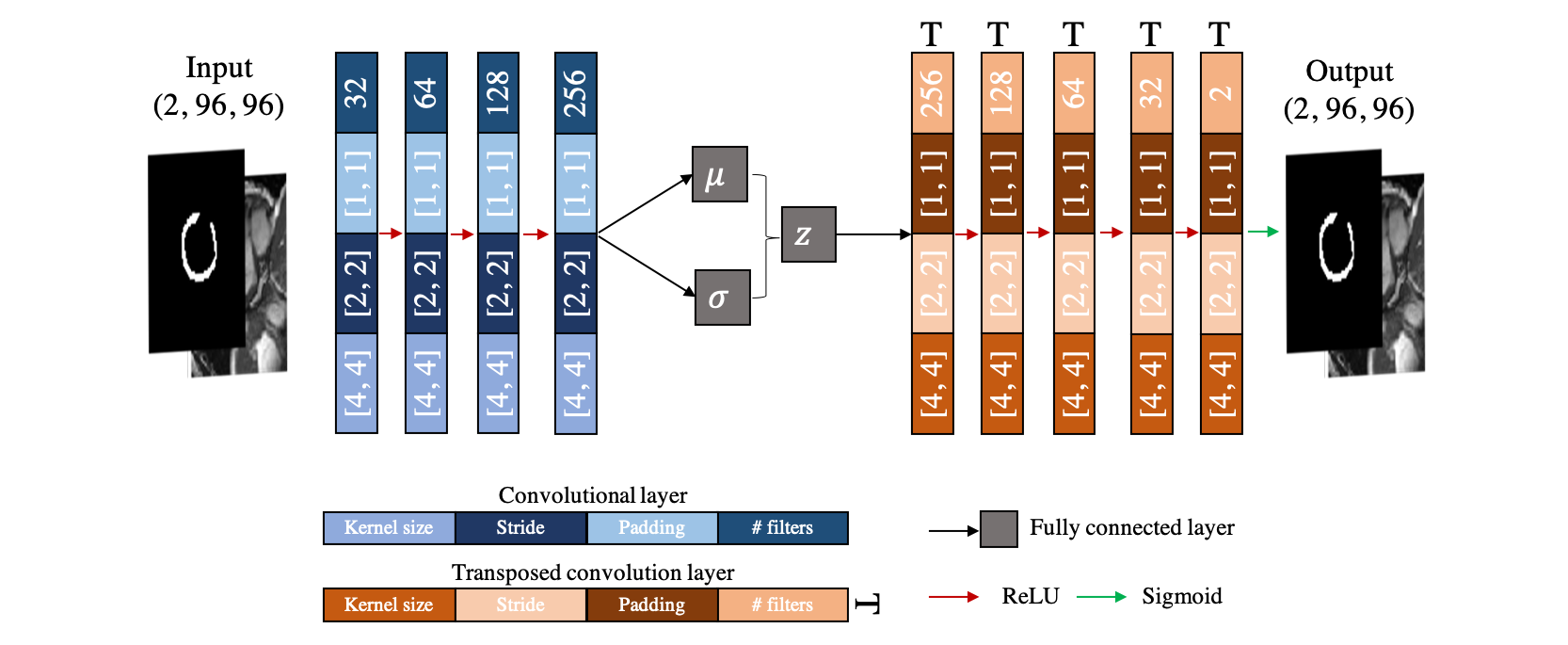}
\caption{Network architecture of the VAE.} \label{figureS1}
\end{figure}

\begin{table}[]
\caption{Ablation study of the latent space dimension $m$ and the hyperparameter $beta$. The mean absolute error (MAE) between predicted and true Dice metrics on UKBB validation set are reported. $m$=8 and $\beta=0.01$ were selected as the optimal parameters according to the results.}

\centering
\begin{tabular}{|c|c|c|c|c|}

\hline
\multirow{2}{*}{\ \ $m \ \ $} & \multicolumn{4}{c|}{$\beta$} \\ \cline{2-5} 
 & 0 & 1E-3 & 1E-2 & 1E-1 \\ \hline
2 & 0.095 & 0.094 & 0.105 & 0.502 \\ \hline
4 & 0.086 & 0.092 & 0.114 & 0.416 \\ \hline
8 & 0.088 & 0.072 & \textbf{0.068} & 0.229 \\ 
\hline
16 & 0.147 & 0.141 & 0.095 & 0.091 \\ \hline
\end{tabular} \label{tableS1}
\end{table}

\end{document}